\documentclass[aps,prl,floats,twocolumn]{revtex4}
\usepackage{bm}
\usepackage{epsfig}
\usepackage{rotating}

\newcommand{\nn}{\nonumber}
\newcommand{\beq}{\begin{equation}}
\newcommand{\eeq}{\end{equation}}
\newcommand{\bea}{\begin{eqnarray}}
\newcommand{\eea}{\end{eqnarray}}
\newcommand{\ben}{\begin{eqnarray*}}
\newcommand{\een}{\end{eqnarray*}}

\newcommand{\lambdaqcd}{\Lambda_{\rm QCD}}

\def\D0{D\O}

\newcommand{\psl}{\not\!p}

\begin{document}

\title{A New QCD Correction to Gauge Boson Decay into Heavy Flavor }

\author{Yu Jia}
\affiliation{Department of Physics and Astronomy, Michigan State University, 
	East Lansing MI 48824}

\date{\today}

\begin{abstract}

We find that, at order $\alpha_s$, the partial width of $Z^0$  to 
heavy flavors receives a power correction from a novel QCD mechanism,
which is not suppressed by inverse powers of $M_Z$, 
but only by two unknown  $O(\lambdaqcd/m)$  constants.
The hadronic $W$ width also receives a similar correction.
These parameters may be fitted from the 
global electroweak analysis, and consequently
the Standard Model predictions of
various electroweak observables will be updated.
This new mechanism is of no help to reconcile the 
discrepancy in $b$ forward-backward asymmetry. 
We also point out the implication of
this mechanism to heavy flavor production 
in other collider experiments.

\end{abstract} 

\maketitle

\vspace{0.33 in} 

The precision electroweak measurements at LEP and SLC
provide the most accurate
knowledge of the fundamental parameters in the Standard Model (SM),
and meanwhile place stringent constraints on 
the new physics beyond SM~\cite{Group:2002mc}.
Though the overall agreement between the measurements and 
the SM fits
is acceptable, few acute discrepancies still persist~\cite{pdg}.
For the $Z$-pole observables, there  are currently a 2.4 $\sigma$ discrepancy
in  the forward-backward asymmetry of $b$ quark ($A_{FB}^b$),
and a 1.9 $\sigma$ deviation in the peak hadronic
cross section ($\sigma^0_{\rm had}$).
While there was a significant deviation in
$R_b\equiv \Gamma[b\bar{b}]/\Gamma_{\rm had}$ in past years, 
the latest measurement is consistent with the SM prediction 
at 1 $\sigma$ level.

In order to  search the possible hint of new physics,
one needs first scrutinize
all the uncertainties within SM.
Ubiquitous (nonperturbative) QCD effects constitute
a particularly important and difficult source of the 
theoretical uncertainties.
This is  best illustrated by muon $g-2$,
where the current theory error is
dominated by the 
low energy
hadronic contribution~\cite{Davier:2002dy}.

In this Letter, we will report a novel QCD mechanism 
which renders a 
new correction to the partial width of $Z^0$ to heavy flavors
at order $\alpha_s$.
It can be calculated in perturbative QCD, up to two
unknown constants of $O(\lambdaqcd/m)$.
Incorporating this new effect will generally influence 
the current SM predictions of 
many electroweak observables, 
especially $\Gamma_{\rm had}$, $\sigma^0_{\rm had}$,
as well as $R_b$, $R_c$.
It will also affect the hadronic $W$ width in a similar way.
We also comment on the implication of this mechanism
to the heavy flavor production in other collider experiments.

The hadronic $Z^0$ width has been measured to per mille 
accuracy.
To match such a precision, radiative corrections from both QCD 
and electroweak sectors must be computed in comparable orders.
For instance, perturbative QCD corrections 
have been included to 3-loop order for 
massless quark, and $b$ quark mass effects have been included up to
order of $m_b^4/M_Z^4$~\cite{Chetyrkin:1996ia}.

Aside from the leading power contribution, 
the hadronic width of $Z^0$ are also affected by power corrections.
For $Z^0$ decaying into light hadrons, 
the well-known
power corrections are characterized by 
the quark and gluon  condensates, whose effects are
suppressed by a factor of
$1/M_Z^4$.  
Therefore they can be  neglected in accordance
with present experimental precision.
 
The situation for heavy flavors is dramatically different.
Because the heavy quark mass sets a new scale, 
the most significant power correction may start at order 
$\lambdaqcd/m$, thus more important  than 
those suppressed by powers of $\lambdaqcd/M_Z$
and $m/M_Z$. 
Surprisingly, 
this possibility has been largely overlooked,
perhaps due to the difficulty for the
standard Operator Product Expansion (OPE) 
to tackle such a multi-scale problem. 

However, as we will see, 
the recently-developed {\it heavy-quark recombination} 
mechanism (HQR)~\cite{Braaten:2001bf,Braaten:2001uu,Braaten:2003vy,Jia:2003ct}
can realize such an $O(\lambdaqcd/m)$ correction
to $Z^0$ hadronic width.
In fact, when HQR was introduced and
first applied to the $B$ hadroproduction,
it was shown that HQR generates an  $O(\lambdaqcd/m_b)$ 
correction to the total $B$ cross section~\cite{Braaten:2001bf}.
With a more concrete hadronization picture, HQR was originally motivated
as a ``higher twist" mechanism,
to supplement the usual heavy quark fragmentation.
However, by studying $Z^0$ decay into $B$ at $O(\alpha_s^2)$,
one recently realizes that under some circumstances,  
HQR can overlap with the fragmentation mechanism~\cite{Jia:2003ct}.
In that work, the $O(\lambdaqcd/m_b)$  contribution from HQR must be
identified with the contribution to the fragmentation function.
Had we been able to extract the finite 
``higher twist" contribution
by removing the ``leading twist" term,
it would represent an $O(\alpha_s^2 \lambdaqcd m_b/M_Z^2)\sim 10^{-6}$
correction to the partial width of $Z^0$ to $b\bar{b}$, 
thus  negligible with current experimental precision.
In fact, the goal of this work is to show there is a new HQR
process occurring at order $\alpha_s$ only, 
with a genuine ``higher twist"
contribution of order $\lambdaqcd/m_b$. Therefore, it is mandatory
to consider its impact on the heavy flavor observables.

The central picture of HQR is that, a heavy quark 
can bind with a light {\it parton} which emerges from 
the hard-scattering
and carries an $O(\lambdaqcd)$ momentum
in the rest frame of heavy quark. 
It is somewhat analogous to that a proton can capture an incident
low energy electron to form a hydrogen atom.
The word {\it parton} deserves some elaboration. 
Intuitively, one expects that $b$ and $\bar q$
are more inclined to bind into a $\overline{B}$ meson,
and $bq$ diquark tends to evolve into a $b$ baryon.
Indeed, the $b\bar{q}$~\cite{Braaten:2001bf} and $cq$~\cite{Braaten:2003vy} 
recombination have been  developed and applied to a variety of 
fixed-target experiments to explain the observed 
charm meson and baryon asymmetries~\cite{Braaten:2001uu,Braaten:2003vy}.

The last recombination mechanism awaiting exploration
then is the {\it $bg$ recombination}, 
when the parton is a gluon.
At first sight,  it seems in contradiction with
the picture of constituent quark model.
Indeed, this objection is justified for the heavy quarkonium case, e.g., 
it is unlikely for $bg$ to evolve into $B_c$.
However, the dynamics of heavy-light hadron is rather 
different from that of heavy quarkonium.
Since the soft gluon in $bg$ can easily split into $q\bar{q}$ pair,
and $bg$ can easily pick up a light antiquark from the vacuum to hadronize, 
there is no dynamical reason for $bg$ recombination
to be suppressed relative to  $b\bar q$ and $b q$ recombination.
On the other hand,  $bg$ recombination does receive 
one parametric suppression from the large $N_c$ consideration. 
There is a $1/N_c$ suppression for $bg$ to evolve into 
a $\overline{B}$ meson
relative to $b\bar{q}$, since it requires an extra $q\bar{q}$ pair.
Similarly, it is even less probable for $bg$  to evolve into a
$b$ baryon than $\overline{B}$, due to a further 
penalty of $1/N_c$.

\begin{figure}[bt]
  \centerline{\epsfysize= 3.1 truecm \epsfbox{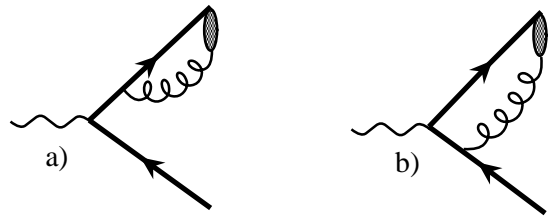}  }
 {\tighten
\caption{Feynman diagrams for the $bg$ recombination process
$ Z^0 \rightarrow bg(n) +\overline{b}$. 
Wavely, solid and curly
lines stand for $Z^0$ boson, $b$ and gluon, respectively.
The shaded blob represents the hadronization of $bg(n)$
into a $\overline{B}$ hadron plus anything.
Note only diagram b) has a nonzero contribution.
}
\label{zbgb} }
\end{figure}

At order $\alpha_s$, bottoms are produced through 
$Z^0 \rightarrow b \bar{b}  g$.
This represents an ordinary 3-jet event if each 
parton independently fragments.
Nevertheless, in a small corner of phase space where $g$ is soft in 
the rest frame of $b$, they can form an intermediate $bg$ 
state with definite color and angular momentum. 
This state then hadronizes into a $b$ hadron 
plus additional soft hadrons.
Therefore, we end up with a jet containing a $b$ hadron from the recombination
and a recoiling $\bar b$. The corresponding Feynman diagrams are
depicted in Fig.~\ref{zbgb}. 
We label the momenta of $bg$ and $\overline{b}$ by $p$ and $\overline{p}$.

The $b\bar q$ recombination
respects a simple multiplicative factorization~\cite{Braaten:2001bf}.
It is suggestive that
the inclusive $\overline{B}$ production  
from $bg$ recombination 
may be also written in a factorized form
(The symbol $\overline B$ schematically 
represents any ground state hadron containing a $b$ quark):
\bea
\label{HQR:factor}
\Gamma [\overline B] = \sum_n  {\hat\Gamma}
[Z^0\rightarrow bg(n) + \overline{b} ]\,
\xi[bg(n) \rightarrow \overline{B}] \,, 
\eea 
where $\hat\Gamma_n$ are the perturbatively calculable partonic cross sections, 
and $\xi_n$ are so-called recombination factors, 
which roughly amount to the probability for $b$ and $g$ 
to evolve into a state including $\overline B$. 
The color and angular momentum quantum numbers of $bg$
are collectively labeled by $n$. 
The inclusive $B$ production  from $\overline{b}g$ 
recombination is identical to Eq.~(\ref{HQR:factor}),
because of  $CP$ invariance.

These $\xi_n$ factors are analogous to those
$\rho_n$, $\eta_n$ associated with $b\bar q$ and $b q$ recombination.
Thus far, a rigorous definition in terms of 
nonperturbative matrix elements is only
available for $\rho_n$~\cite{Chang:2003ag},
but generalization to $\eta_n$ and $\xi_n$ 
should be straightforward. 
We also assume $\xi^b_n \sim \lambdaqcd/m_b$, 
the same as $\rho_n$~\cite{Braaten:2001bf,Chang:2003ag}. 

The $bg$ 
that emerges from the hard-scattering
can be in either of three irreducible color representations:
$3$, $\overline{6}$ or $15$. 
Fortunately,
due to very simple color structure, 
only color-triplet channel survives in this process. 
The color-triplet state $bg$ can be labeled by a 
fundamental $SU(3)$ index $i$ and has the 
normalized color wave function
\bea
| bg(3)_i \rangle &=& 
{\sqrt{3} \over 2} \, T_{a\:i}^{\;\,j}\,|b_j\rangle\,|g_a\rangle\,,
\label{proj-3}
\eea       
where $SU(3)$ generators $T_a$ $(a=1\cdots 8)$ 
satisfy the normalization 
${\rm tr}(T_aT_b)=\delta_{ab}/2$.
In more complicated  $bg$ recombination processes, 
the other two color channels will also contribute.

We also need project the $bg$ onto the states of  definite 
angular momentum.
We will consider only the S-wave states, 
because the higher orbital angular momentum states will 
result in  towers of $\lambdaqcd/m_b$ suppression.
Therefore, the $bg$ state can be either $^2S_{1/2}$ or $^4S_{3/2}$,
corresponding to a spin-$1\over 2$  or 
spin-$3\over 2$ (Rarita-Schwinger) fermion.
The momentum eigenstates of these fermions 
are conveniently represented by 
a vector-spinor $u_J^\mu(p)$~\cite{Falk:1991nq}, 
which carries an extra Lorentz index relative to the Dirac spinor. 
Both of $u_{1/2}^\mu(p)$ and $u_{3/2}^\mu(p)$ satisfy the 
on-shell condition $\psl\,u_J^\mu(p)=m_b$ and transversality
constraint $p\cdot u_J(p)=0$. 
The spin-$3\over 2$  vector-spinor 
is subject to a further constraint $\gamma_\mu u_{3/2}^\mu=0$.

In addition to projecting the amplitude 
onto states of definite color and spin,
we need isolate those most singular terms when the 
gluon has small momentum in the rest frame of the $b$
quark~\cite{Braaten:2001bf}.
This can be accomplished by setting $p_g \to x_g p$, where $x_g$ is a
momentum fraction that is presumably of order $\lambdaqcd/m_b$,
and then taking the limit $x_g \to 0$.  
The factor $1/x_g$ appearing in the amplitude must be absorbed into
the nonperturbative  factor $\xi_n$.
Pragmatically, this can be described by a simple prescription. 
First we  make the following substitution in the $b\bar{b}g$ amplitude:
\beq
  \overline u_j(p_b) \, \epsilon^{*a\mu}(p_g) \rightarrow 
  x_g \,\xi[bg(3, J)]^{1\over 2}\,{\sqrt{3}\over 2} \,T_{a\,j}^{\;\;i} 
  \, m_b\, \overline u_J^\mu(p)\,,
\eeq
then set $p_g=x_g\,p$  in the rest of the amplitude and take the limit
$x_g\rightarrow 0$.

To justify a perturbative treatment, 
we need show the parton process is really governed by 
the short distance scale. 
The characteristic momentum carried by the recombining gluon is
$p_g\sim\lambdaqcd v$, 
where $v=p/m_b$ is the 4-velocity of $bg$.
Therefore, the virtual $b$ quark in Fig.~\ref{zbgb}a) travels a distance of 
order $\sqrt{\lambdaqcd \over m_b}/\lambdaqcd$
before turning on-shell, while the corresponding distance in Fig.~\ref{zbgb}b)
is of order $\sqrt{m_b\,\lambdaqcd \over M_Z^2}/\lambdaqcd$, 
both are much smaller than the confinement length $1/\lambdaqcd$.

Nevertheless, Fig.~\ref{zbgb}a) is potentially worrisome, since it is
reminiscent of 
a $\delta(1-z)$ like fragmentation contribution, 
of which we need get rid, to avoid double counting.
The subamplitude of Fig.~\ref{zbgb}a) involves a term 
\bea
\label{figa:vanish}
   \overline u_J^\mu(p)\gamma_\mu (\psl+m) &=&
   2\,\overline u_J(p) \cdot p- 
   \overline u_J^\mu(p)(\psl-m)\gamma_\mu.
\eea
Pleasingly, by transversality and the equation of motion,
one immediately sees that it vanishes for both $^2S_{1/2}$ and $^4S_{3/2}$
states. 
So only Fig.~\ref{zbgb}b) needs to be included,
which has nothing to do with  the $b$ fragmentation, 
since the recombining gluon is emitted from  $\bar{b}$.
Therefore it represents a genuine ``higher twist" contribution.

To be specific, let us first consider the parton cross section for 
the $^2S_{1/2}$ state.
After some simple manipulation similar to  Eq.~(\ref{figa:vanish}), 
we can reduce the amplitude into 
\bea
\label{ampl:fig:b}
  \hat{\cal M} \left[ bg(3, {^2S_{1\over 2}}) \right ]&\propto& 
    {m_b \over p\cdot \overline{p} } \,
  \overline u_{1\over 2}(p)\cdot \overline{p} \,\Gamma^{\alpha} 
  v(\overline p)\,\epsilon_{\alpha}(Z) \,,
\eea
where $\Gamma^\alpha$ is the $Zb\bar{b}$ coupling,
and $\epsilon_{\alpha}(Z)$ is the polarization vector of $Z^0$.
Squaring the amplitude and
summing over the color are standard.
Less familiar is the sum over two polarizations of $u^\mu_{1/2}$, 
yet can be accomplished
with the aid of the following formula~\cite{Falk:1991nq}:
\bea
\label{spinhalf:sumpolar}
\sum u^\mu_{1\over 2}(p) \,\overline{u}^\nu_{1\over 2}(p) \,
\overline{p}_\mu \, \overline{p}_\nu
&=& {1\over 3} (\psl+m) [(\overline{p} \cdot v)^2 -\overline{p}^2]\,.
\eea
Note this resembles summing over polarizations
of a Dirac spinor, up to a normalization.
Finally we arrive at a succinct expression:
\bea
\label{spinhalf:dw}
\hat \Gamma [ bg(3, ^2\!S_{1/2})]  &=&
{16\,\pi\,\alpha_s(M_Z) \over 9} \,{1-4\gamma \over (1-2\gamma)^2}\,
\Gamma_0[b\bar{b}] \,,
\eea
where $\gamma=m_b^2/M_Z^2$,
and $\Gamma_0$ is the lowest order $Z^0$ partial decay
width into $b\overline{b}$:
\bea
\Gamma_0[b\bar{b}] &=& 
{G_F\,M_Z^3 \over 2\,\sqrt{2}\,\pi}\, 
\beta \,
\left [ g_V^{b\,2}(1+2\gamma)
+ g_A^{b\,2}(1-4\gamma) \right ],
\eea       
where $\beta=\sqrt{1-4\,\gamma}$, 
the vector coupling $g_V^b=-{1\over 2}+{2\over 3}\sin^2 \theta_W$ and
the axial-vector coupling  $g_A^b=-{1\over 2}$.

The HQR parton cross section for the $^4\!S_{3/2}$ state
can be derived in a similar way. 
It turns out $\hat \Gamma [ bg(3, ^4\!S_{3/2})]=
2\,\hat \Gamma [ bg(3, ^2\!S_{1/2})]$, 
as expected from counting the polarizations.
It is remarkable that they
are simply $\Gamma_0$ multiplying a constant. 
Of course, this simplification is mostly due to
the vanishing of Fig.~\ref{zbgb}a).

The kinematic factor appearing in Eq.~(\ref{spinhalf:dw})
suppresses the HQR contribution
when the center-of-mass energy is near the
$b\bar{b}$ threshold. 
Nevertheless, on the top of $Z$ pole,
setting this kinematic factor to 1
is an excellent approximation for $b$ and $c$.
Since the HQR parton cross sections
are finite when $\gamma = 0$,
it might be tempting to apply Eq.~(\ref{spinhalf:dw}) also to 
the light hadron production.
However, the validity of HQR factorization in Eq.~(\ref{HQR:factor}) 
is crucially based on the heavy quark dynamics,  
thus one should not erroneously
use Eq.~(\ref{spinhalf:dw}) to describe $Z^0$ decay into
light hadrons, by literally taking $\gamma\to 0$.

It may look surprising that the HQR parton cross sections are $O(1)$
relative to $\Gamma_0$, not suppressed by $m_b^2/M_Z^2$ 
as indicated in Eq.~(\ref{ampl:fig:b}).
In fact, the explicit suppression from $m_b$ 
in Eq.~(\ref{ampl:fig:b}) is compensated by $\overline{p}\cdot v$
in Eq.~(\ref{spinhalf:sumpolar}).
This non-suppression has important implication to the heavy 
flavor production in the hadron collision. 
By examining the $bg$ recombination process 
$gg\rightarrow bg(3)+\overline{b}$, 
one finds that the differential cross section 
$d\hat{\sigma}[bg(3)]/dp_t^2\sim 1/p_t^4$
at $p_t\gg m_b$,
{\it a priori} of the same order as fragmentation.
In fact, it is equal to the differential cross section for
$gg$ annihilating into massless $q \bar{q}$ pair,
up to a numerical constant.
This finding diametrically challenges the general belief, that
any non-fragmentation process at large $p_t$
will be suppressed by powers of $1/p_t^2$.

To expedite the analysis, we choose the following linear combination
of recombination factors:
$\xi_{3 \rm eff} =\xi[bg(3, ^2\!S_{1/2})] + 
2\,\xi[bg(3, ^4\!S_{3/2})]$.
We further define
$\xi_{3 \rm tot}  = \sum_{\overline B} \xi_{3\rm eff}[bg\to {\overline B}]$, 
where the sum is over all the lowest-lying $b$ mesons and baryons.
So the net contribution of this mechanism to the partial width of 
$Z^0$ to $b\bar{b}$  is
\bea
\label{bwidth:corr}
\Delta \Gamma [b\bar{b}]  &=&
{32\,\pi\,\alpha_s \,\xi^b_{3 \rm tot}\over 9} \,\Gamma_0[b\bar{b}] \,,
\eea
where we have doubled Eq.~(\ref{spinhalf:dw}) to include the 
contribution from  $CP$ conjugate of Eq.~(\ref{HQR:factor}).
This is the crux of this work. 
From now on, we will use 
$\xi_3$ as a shorthand for $\xi_{3 \rm tot}$ for simplicity.

The correction to the partial width of $Z^0$ into charm 
can be obtained by simply replacing $\xi_3^b$ and $\Gamma_0[b\bar b]$
with $\xi_3^c$ and $\Gamma_0[c\bar{c}]$ in Eq.~(\ref{bwidth:corr}).
The heavy quark symmetry suggests that
$\xi_3^c= \xi_3^b\, m_b/m_c \approx 3 \,\xi_3^b$.
However, a caveat is that,
the symmetry-breaking effect may be large
in the real world and practically we better 
treat them as two independent unknown parameters.

The $bg$ recombination can also make a correction to
the hadronic width of $W$ boson, through the decay channels 
$W^+\to \bar{b}g+ c(u)$ and $W^+\to cg+\bar{s}(\bar{d})$
(Same reasoning can be also applied to $t\to bg+ W^+$).
Note it is not necessary for the recoiling quark to be heavy.
The calculations are essentially the same as described above for 
$Z^0$ decay.
However, unlike $\Gamma_Z$ which is measured to per mille accuracy,
$\Gamma_W$ is measured with a $3\%$ error. 
Therefore the $Z$-pole observables are more sensitive probes to 
ascertain the effects of this mechanism.

\begin{table}[tb]
\caption{
Some selected electroweak variables, where
$R_e \equiv \Gamma_{\rm had}/\Gamma[e^+e^-]$ and
$\sigma^0_{\rm had} \equiv 
12\pi \Gamma[e^+e^-]\Gamma_{\rm had}/(M_Z^2\Gamma_Z^2)$. 
Listed are the latest measurements~\cite{Group:2002mc} 
and the SM predictions~\cite{pdg}. 
Pull is defined as $(O_{\rm meas}- O_{\rm fit})/ \sigma_{\rm meas}$.}
\label{zpole} 
\begin{center}
\begin{tabular}{ccc|ccc|ccc|ccc}
&    && & Meas. & && Predictions &  && Pull &
 \\ \hline 
& $\Gamma_{\rm had} $ [GeV] &&  &$1.7444\pm0.0020$&  && $1.7429\pm 0.0015$ &
&&--&\\
& $\Gamma_Z $ [GeV]  &&  &$2.4952\pm0.0023$&  && $2.4972\pm0.0011$ & && -0.9&\\
& $\sigma^0_{\rm had}$ [nb] &&  &$41.541\pm0.037$&  && $41.470\pm 0.010$ & &&
1.9 &\\
& $R_e$  && & $20.804\pm0.050$  & && $20.753\pm 0.012$ & && 1.0&\\ 
& $R_b$  &&  &$0.21644\pm0.00065$&  && $0.21572\pm 0.00015$ & && 1.1&\\ 
& $R_c$  &&  &$0.1718\pm0.0031$&  && $0.17231\pm 0.00006$ & && -0.2 &\\ \hline
& $\Gamma_W$ [GeV] &&  &$2.139\pm0.069$&  &&  $2.0921\pm0.0025$ & && 0.7& 
\end{tabular}
\end{center}
\end{table}

The discrepancy in $A_{FB}^b$ is believed to either 
be an experimental problem, or originate 
from some tree-level new physics effects 
on $Zb\bar{b}$ vertex~\cite{pdg}.
Our $bg$ recombination doesn't have much impact on
the forward-backward asymmetries of $b$ and $c$,
since the role of this mechanism is simply rescaling both
forward and backward cross sections by a common factor,
which cancels out in the ratio.

From Table~\ref{zpole}, we can see the general agreement 
between the measurements and
SM fits is rather good, 
therefore demanding  $\xi_3$ must be small.
Naively, 
if we assume including the corrections 
in Eq.~(\ref{bwidth:corr})  
doesn't affect the values of SM fit parameters,
then the relative variations of the SM
predictions listed in Table~\ref{zpole} are
(For simplicity, we take $\xi_3^c = 3 \,\xi_3^b$ temporarily):
\bea
   \delta \Gamma_{\rm had} / \Gamma_{\rm had} \approx  
   \tau\,
   (R_b+3 R_c) &\approx&  \xi_3^b, 
\nn \\  
   \delta \Gamma_Z / \Gamma_Z  =  (\Gamma_{\rm had}/\Gamma_Z)\,
   \delta \Gamma_{\rm had} / \Gamma_{\rm had} 
   &\approx&  0.7 \, \xi_3^b, 
\nn  \\
   \delta \sigma^0_{\rm had} / \sigma^0_{\rm had} =  
  (1-2 \Gamma_{\rm had}/\Gamma_Z )\, \delta \Gamma_{\rm had} / \Gamma_{\rm had}
   &\approx& -0.4 \, \xi_3^b,
\nn  \\
   \delta R_e / R_e =   \delta \Gamma_{\rm had} / \Gamma_{\rm had}
    &\approx&  \xi_3^b,
\nn \\
  \delta R_b / R_b \approx  \tau\,
   (1-R_b-3 R_c) &\approx&  0.4\, \xi_3^b,
\nn \\
   \delta R_c / R_c \approx  \tau\,
   (3-R_b-3 R_c) &\approx&  3 \,\xi_3^b,
\nn  \\
   \delta \Gamma_W / \Gamma_W \approx  
   3\,\tau/2\, (0.707\:{\rm GeV}/\Gamma_W) 
   &\approx& 0.7\, \xi_3^b, 
\nn
\eea
where $\tau\equiv 32 \pi \alpha_s \,\xi_3^b/9$, 
and $\alpha_s(M_Z)=0.12$ is used.
In the last row, we use 
$\Gamma[W^+\to u_i\bar{d}_j]\approx  0.707\,|V_{ij}|^2\:{\rm GeV}$~\cite{pdg},
and only include the Cabbibo-favored channels 
$W^+\to cg+ \bar{s}(\bar{d})$.
To be compatible with  most variables,
$\xi_3^b \sim 10^{-3}$ looks reasonable. 
It will drive $R_e$, $R_b$ and $\Gamma_W$ towards the correct direction,
but deteriorate $\Gamma_Z$, $\sigma^0_{\rm had}$ and $R_c$.
Note the correction to $R_c$ is much more significant  than to $R_b$.
A $\xi_3^b$  as large as 0.01 becomes apparently unacceptable,
since it would bring up the deviations in $\sigma^0_{\rm had}$ and $R_c$ 
to 6.4 $\sigma$~\footnote{ 
Perhaps the $\sigma^0_{\rm had}$ discrepancy is not so glaring. 
It was recently suggested~\cite{deBoer:2003xm} that,
the luminosity in four LEP experiments may have been underestimated, 
and correcting  that will decrease the measured $\sigma^0_{\rm had}$ value.}
and 1.8 $\sigma$, respectively.

Since many quantities are correlated in a complicated way,
an unbiased strategy is to incorporate this mechanism into 
the global electroweak analysis.
Consequently, the recombination factors $\xi_3^b$ and $\xi_3^c$,
can be fitted together with the SM parameters
$M_Z$, $M_H$, $m_t$, $\alpha_s(M_Z)$
and $\Delta\alpha^{(5)}_{\rm had}$.
It will be very interesting to see how these parameters vary,
and how much the quality of 
global electroweak fit may improve.

To summarize, we have studied a new QCD mechanism 
that generates power corrections to partial widths of
$Z^0$ and $W$ at $O(\alpha_s)$, 
which have previously eluded the OPE analysis.
The present precision of electroweak measurements 
requires these terms must be included.

The phenomenological consequences of this mechanism 
to other collider experiments should be investigated.
For example, we have noticed one unusual feature of 
$bg$ recombination in hadron collision --
it is formally of the same order as 
fragmentation contribution at large $p_t$. 
This is in sharp contrast to $b\bar{q}$ and $bq$ recombination, 
whose effects  
are suppressed by $\lambdaqcd m/p_t^2$~\cite{Braaten:2001bf}.
In addition, because the dominant contribution to
heavy flavor hadroproduction arises from
the gluon-initiated  subprocess, 
we expect $bg$ recombination is more important 
than $b\bar{q}$ and $b q$ recombination.
 
Unfortunately, 
a rather (unnaturally) small $\xi_3$ inferred 
from this work
makes the color-triplet contribution 
virtually invisible in hadron and $ep$ collider experiments,
due to much cruder consistency between
data and the NLO QCD predictions.
Since $\xi_6$ and $\xi_{15}$ remain unconstrained,
we hope they perhaps are much larger than $\xi_3$,
so that the color-$\overline{6}$ and $15$  channels 
may result in noticeable effects.

I thank C.-P.~Yuan for discussing
electroweak observables.
This work is supported by the
National Science Foundation under Grant No.~PHY-0100677.


\begin{references}

\bibitem{Group:2002mc}
LEP Electroweak Working Group and SLD Heavy Flavour Group,
hep-ex/0212036; The latest measurements can also be found
in http://lepewwg.web.cern.ch/LEPEWWG. 

\bibitem{pdg}
See the review by J.~Erler and P.~Langacker, 
Phys.\ Rev.\ D {\bf 66}, 010001-98 (2002) 
[Particle Data Group (K.~Hagiwara {\it et al.})];
J.~Erler,
hep-ph/0212272.

\bibitem{Davier:2002dy}
M.~Davier {\it et al.},
Eur.\ Phys.\ J.\ C {\bf 27}, 497 (2003).



\bibitem{Chetyrkin:1996ia}
K.~G.~Chetyrkin, J.~H.~Kuhn and A.~Kwiatkowski,
Phys.\ Rept.\  {\bf 277}, 189 (1996) and references therein.


\bibitem{Braaten:2001bf}
E.~Braaten, Y.~Jia and T.~Mehen,
Phys.\ Rev.\ D {\bf 66}, 034003 (2002).


\bibitem{Braaten:2001uu}
E.~Braaten, Y.~Jia and T.~Mehen,
Phys.\ Rev.\ D {\bf 66}, 014003 (2002);
Phys.\ Rev.\ Lett.\  {\bf 89}, 122002 (2002).


\bibitem{Braaten:2003vy}
E.~Braaten {\it et al.}, hep-ph/0304280.


\bibitem{Jia:2003ct}
Y.~Jia, hep-ph/0305172.


\bibitem{Chang:2003ag}
C.~H.~Chang, J.~P.~Ma and Z.~G.~Si, 
hep-ph/0301253.


\bibitem{Falk:1991nq}
A.~F.~Falk,
Nucl.\ Phys.\ B {\bf 378}, 79 (1992).

\bibitem{deBoer:2003xm}
W.~de Boer and C.~Sander,
hep-ph/0307049.

\end{references}
\end{document}